\documentclass[final,numberedheadings]{aipproc}

\layoutstyle{6x9}

\usepackage{amsmath,amssymb,amsfonts,amsthm,amscd}
\usepackage[mathscr]{eucal}
\usepackage{epsfig}
 
\newcommand{\plaq}[5]{

\setlength{\unitlength}{.5in}

\begin{picture}(3.25,2)(0.5,1)

\put(.9,2){$\blacktriangle$}

\put(0.55,2){$#4$}

\put(1,1){\line(0,1){2}}

\put(1,1){\line(1,0){2}}

\put(2,.9){$\blacktriangleright$ }

\put(1.9,0.7){$#1$}

\put(1,3){\line(1,0){2}}

\put(2,2.9){$\blacktriangleright$ }

\put(1.9,3.15){$#3$}

\put(2.9,2){$\blacktriangle$}

\put(3.15,2){$#2$}

\put(3,1){\line(0,1){2}}

\put(2,2){$#5$}

\end{picture}
}

\theoremstyle{definition}

\begin{document}

\title{Path space forms and surface holonomy}


\author{Saikat Chatterjee}
{address = {S.~N.~Bose National Centre for Basic Sciences, Block JD,
 Sector III, Salt Lake, Kolkata 700098, West Bengal, INDIA    } }
\author{Amitabha Lahiri}
{address = {S.~N.~Bose National Centre for Basic Sciences, Block JD,
 Sector III, Salt Lake, Kolkata 700098, West Bengal, INDIA    } }
\author{Ambar N. Sengupta}{address = {Department of Mathematics,
 Louisiana State University  Baton
Rouge, Louisiana 70803, USA   } }

\begin{abstract}
  We develop parallel transport on path spaces from a differential
  geometric approach, whose integral version connects with the
  category theoretic approach. In the framework of 2-connections,
  our approach leads to further development of higher gauge theory,
  where end points of the path need not be fixed.\thanks{Talk
    delivered by S. Chatterjee at XXVIII WGMP, 28th June-4th July,
2009. Bialowieza, Poland.}
    
\end{abstract}  

\maketitle

\section{Introduction}
The issue of parallel transport along surfaces has been the subject of
a growing body of literature (\cite{catta, baez, baezschreiber, girelli},
to mention a few). The importance of surface parallel transport lies
in the fact that to describe string like objects, it is the natural 
framework. Ordinary gauge theory describes the interactions
between particles, where the gauge group is the relevant symmetry
group of the particles; if instead of particles we have string-like
objects, a higher gauge theoretic structure becomes more natural, and category
theoretic argument show that, unless the group is Abelian, a
single group is not sufficient for that purpose. In this article we
describe a bridge between the differential geometric and category
theoretic approaches to the problem of surface parallel
transport. We start with   path-space forms to  build up the
necessary frame work. Then starting from a differential geometric
approach we develop the category theoretic structure of the
parallel transport on path space. We work in a principal fiber
bundle $(\Pi, P, M)$, with the gauge group $G$. We consider the
$\bar A$-horizontal path space ${\mathcal P}_{\bar A} P$, where
$\bar A$ is a $LG$ valued connection on $P$. Keeping in mind that a
single group is insufficient to describe   surface parallel
transport, we introduce another group $H$ related to $G$ and
construct a connection $\mathcal A$ for the path space from another
$LG$ valued connection $A$ and a $LH$ valued $2$-form $B$. With the
help of this connection we develop   parallel transport on the
path space, which leads to a natural construction of an integrated
picture or category structure for surface parallel transport.\\

\medskip

\underline {\textit {Path space forms}} \\

We start with the construction of `path-space forms' on a  
manifold $M$. Let us first define the path space ${\cal P}M$ of a
manifold $M$ as the space of all smooth paths in $M$, i.e. if $\gamma \in
{\cal{P}} M$, then $\gamma(t)\in M$ and
\begin{equation}
\gamma:I\rightarrow M \hskip 2.0cm  I=[0,1]\nonumber
\end{equation}
is smooth.
We define the tangent space of the path space as follows: for a
$\gamma \in {\cal{P}}M$, a vector $X \in T_{\gamma}({\cal{P}}M)$ is
given by a vector field $X(t) \in T_{\gamma (t)}(M)$
\cite{catta}. Note here that ${\mathcal{P}}M$ is infinite dimensional,
which is consistent with the fact that to define a `single' vector
on a point $\gamma$ of ${\cal{P}}M$, we need to define a vector field
along a path $\gamma(t)\in M$.

Let  $ev$ be the general evaluation map. i.e.
\begin{equation}
ev:{\cal{P}} M \times I\longrightarrow M,\hskip 0.5cm
ev_{t}\stackrel{\rm def}{=}ev(\cdot,t)  \label{pathspace
  evaluation} 
\end{equation}
Then $ev_{t} : {\cal{P}} M \rightarrow M$ defines $\gamma \mapsto
\gamma(t)$, and given a $p-$form field $\alpha_{p} \in \Omega^{p}M$
we can construct a $p$-form $ev_t^{*}\alpha_{p}$ on the ${\cal{P}}
M$ by pulling it back. The above definition of the path space
tangent vector leads to the following contraction formula
\begin{equation}
ev_{t}^{*}(\alpha_{p})({X_1}, {X_2},\ldots,{X_p})=(\alpha_{p})({X_1(t)},{X_2(t)},\ldots,{X_p(t)}) \label
{path space action} 
\end{equation}
Another  construction of path space forms follows  the method of
K.T. Chen\cite{chen}, \cite{chen1},  known as the `Chen
integral'. We will not  discuss the Chen integral in detail,  only
a first order Chen integral is sufficient for our present
purpose. Higher Chen integrals can  be defined by the method of
iteration. Given $\alpha_{p+1}\in \Omega^{p+1}M$, a first order
Chen integral is defined as 
\begin{equation}
\int_{chen} \alpha_{p+1}\stackrel{\rm
  def}{=}\int_{I}ev^{*}\alpha_{p+1}=\int_{0}^{1} \alpha_{p+1}({\dot
  {\gamma}}(t),\ldots)dt \in \Omega^{p}{{\cal{P}} M} \label{chen} 
\end{equation}
Here $\gamma$ is a path on $M$ defined on the interval
$I=[0,1]$. More explicitly, the contraction formula for a first order Chen
integral reads as 
\begin{equation}
\int_{chen}
\alpha_{p+1}\mid_{\gamma}({X_1},{X_2},\cdot,\cdot,{X_p})=\int_{0}^{1}
\alpha_{p+1}({\dot
  {\gamma}}(t),{X_1}(t),{X_2}(t),\cdot,\cdot,{X_p}(t))dt 
\label{chencontraction}  
\end{equation}
For simplicity, we will often denote
$\int_{I}ev^{*}\alpha_{p}$ as $\int \alpha_{p}$, which should not be
confused with the ordinary integral of $\alpha_{p}$ on a
manifold.\\

\medskip

\underline {\textit {$\bar A$-horizontal path space }}\\

Let us consider a principal $G$-bundle $(\Pi, P, M)$
\begin{equation}
\Pi : P \rightarrow M   \nonumber
\end{equation} 
with the usual right action of the Lie group $G$ on $P$
\begin{equation}
P \times G \rightarrow P: (p,g) \rightarrow pg  \nonumber
\end{equation} 
If $\bar A$ is a connection on this bundle, we consider the
space of $\bar A$-horizontal paths in $P$. An $\bar A$-horizontal
lift  ${\tilde \gamma}$ of a path $\gamma: I \rightarrow M$  satisfies
\begin{equation}
\Pi({\tilde \gamma}(t))= \gamma (t) \nonumber
\end{equation}
This $\bar A$ horizontal path space ${\cal{P}}_{\bar A} P$ can be
viewed as a principal $G$-bundle over ${\cal{P}}M$, for details see
\cite{catta}. It can be shown (Proposition~2.1 in
\cite{saikat}) that if ${\tilde \Gamma} : [0,1]\times [0,1]
\rightarrow P:(t,s)\rightarrow {\tilde \Gamma} (t,s)= {\tilde
  \Gamma}_{s}(t)$ is a smooth map and ${\tilde
  X}_s(t)=\partial_{s}{\tilde \Gamma}(t,s)$,  then {\textit {each transverse path
    ${\tilde \Gamma}_{s}: [0, 1]\rightarrow P$ is $\bar
    A$-horizontal}} implies that {\textit {the initial path $
    {\tilde \Gamma}_{0}$ is $\bar A$-horizontal, and the tangency
    condition}} 
\begin{equation}
\begin{frac}{\partial{\bar A}({\tilde
X}_s(t))}{\partial t}\end{frac}=
F^{\bar A}\left({\partial_t}{\tilde\Gamma}(t,s), 
{\tilde X}_s(t)\right) \label{tngtcond}
\end{equation}
holds. In integral form this is:
\begin{equation}
ev_T^{*}{\bar A}-ev_0^{*}{\bar A} = \int_{0}^{T} F^{\bar
  A} \label{integratedtngtcond} 
\end{equation}
The right hand side is a Chen integral over the interval $[0,T]$. Now we
can define the tangent space $T_{{\tilde \gamma}}{\cal{P}}_{\bar A}
P$  at a point $\tilde \gamma$ of  ${\cal{P}}_{\bar A} P$ as the
space of all vector fields $t \rightarrow {\tilde X}(t) \in
T_{{\tilde \gamma}(t)}P$ along ${\tilde \gamma}$ for which
\eqref{tngtcond} holds, i.e.
\begin{equation}
\begin{frac}{\partial{\bar A}({\tilde
X}(t))}{\partial t}\end{frac}=
F^{\bar A}\left({{\tilde\gamma} '}(t), 
{\tilde X}(t)\right) \label{horizontaltngtvectorcond}
\end{equation}
for all $t \in [0, 1]$. The vertical subspace of  $T_{{\tilde
    \gamma}}{\cal{P}}_{\bar A} P$ 
is the linear space of all vectors ${\tilde X}$ for which ${\tilde
  X}(t)$ is vertical (a more  detailed discussion can be found in
\cite{saikat}).\\

\medskip


\underline {\textit {Parallel transport on  path space}}\\

\begin{figure}[ht]
\setlength{\unitlength}{.5in}
\begin{picture}(6.5,6)(2,1)
\put(2,3){\plaq{}{  }{}{}{H''}}
\put(4,3){\plaq{}{}{}{}{H'''}}
\put(2,1){\plaq{}{  }{}{}{H}}
\put(4,1){\plaq{}{}{}{}{H'}}
\end{picture}
   \caption{No Go Theorem}
   \label{fig:window}
\end{figure}
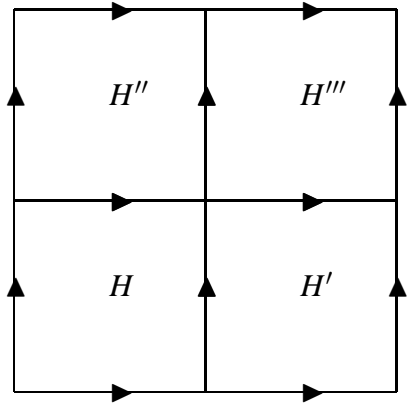

A description of parallel transport on path space by naively
using a connection, with values in the Lie algebra $LG$, on the path space (or
horizontal path space) leads to a serious inconsistency with some natural
requirements. It is natural to require that 'vertical' and
`horizontal' composition of surface parallel transports satisfy a consistency condition:
\begin{equation}
  (H'\bullet H )\times (H'''\bullet H'')=(H' \times
  H''')\bullet (H \times H'') \label{inconsistency} 
\end{equation}
here $H,H',H'',H'''$ are `surface parallel
transport' operators in Figure \ref{fig:window} and $\times$ and
 $\bullet$ denote vertical and horizontal composition for surfaces 
respectively. 

>From this it is clear that if we take the surface parallel transport
operator to be a group element and assign the same composition law
(the group product) for both horizontal and vertical compositions,
the group must be Abelian.  Thus there is a `no-go theorem' (see \cite{baez} for
much more on this).   This problem is avoided by using two groups $G$ and $H$ to
describe surface
parallel transport, and defining different composition laws for the
`horizontal' and `vertical' compositions. The proper  framework  here is
 provided by the notion of a {\textit {Lie $2$-group}}
\cite{baezschreiber1, saikat}, which we discuss below.

A Lie $2$-group is described by two Lie groups $G$ and $H$, along
with a smooth homomorphism $\tau: H \rightarrow G$ and a smooth map 
for $g \in G$ and $h, h' \in H$
\begin{equation}
G \times H \rightarrow H: (g, h)\mapsto \alpha(g)h \nonumber
\end{equation}
where
$\alpha (g)$ is an automorphism of $H$, and the following identities
hold 
\begin{equation}\label{E:taualpha}
\tau (\alpha (g)h)= g \tau(h) g^{-1}, \quad
\alpha(\tau (h))h'=h h' h^{-1}
\end{equation}
(There are fancier, category-theoretic formulations of the notion of Lie $2$-group.)
For simplicity we will denote the derivative mappings $\alpha'(e): LG
\rightarrow LH$ and $\tau'(e):LH \rightarrow LG$ as $\alpha$ and
$\tau$ respectively, here $LG$ and $LH$ are Lie algebras of $G$ and
$H$ respectively.\\

\medskip

\underline{\em Connection form on  path space}\\

Suppose we have a connection $A$ on the bundle $P$ and an $LH$
valued $\alpha$-equivariant (under the right action of $G$)
$2-$form $B$ on $P$, which vanishes on vertical vectors. i.e.
\begin{equation*}\begin{split}
B(X, Y)&= 0, \quad {\hbox{if $X$ or $Y$ is vertical}} 
\\ 
 R_{g}^{*} B &= \alpha(g^{-1})B \quad {\hbox{for all $g \in
    G$}}\end{split}
\end{equation*}
here $R_{g}: P \rightarrow P:p\mapsto pg$ and according to our
convention $\alpha(g^{-1})B=d\alpha(g^{-1})|_{e}B$.

Keeping the `no-go' theorem  in mind,  we
define our connection as
\begin{equation}
{\mathcal A}=ev_1^{*}A + \tau {\int_{0}^{1}B} 
\label{pathspaceconnection} 
\end{equation}

The integration on the right hand side is a first order Chen
integral. For a proof that the right hand side of the
\eqref{pathspaceconnection} is a connection see \cite{saikat}. At the
infinitesimal level, the parallel transport of a path by the
connection ${\mathcal A}$ is equivalent to lifting a given vector
field $X: I \rightarrow TM$, along $\gamma \in {\mathcal P}M$, to a
vector field $\tilde X$ along ${\tilde \gamma}$ such that it is
${\mathcal A}$ horizontal and satisfies the condition
\eqref{horizontaltngtvectorcond} :
\begin{equation}\label{pathtransport}\begin{split}
 A({\tilde X}) + \tau {\int_{0}^{1}B ({\tilde \gamma}'(t),{\tilde
    X}(t))}dt &= 0\\ 
 \begin{frac}{\partial{\bar A}({\tilde
X}(t))}{\partial t}\end{frac} &=
F^{\bar A}\left({{\tilde \gamma} '}(t), 
{\tilde X}(t)\right) \end{split}\end{equation}
Now decomposing a lifted vector ${\tilde X} (t)={\tilde X}_{\bar
  A}^h (t) +{\tilde X}^V (t) $ into horizontal and vertical parts
with respect to the connection ${\bar A}$ and noting that $B$ is
zero on the vertical vectors, it can be shown \cite{saikat} that we
can find a vector field ${\tilde X} (t)$ which satisfies  
\eqref{pathtransport}. The basic
idea in our construction is that the equations
\eqref{pathtransport}   specify `parallel transport' of the
`right endpoint' ${\tilde \gamma}(1)$ and then
\eqref{pathtransport} specifies the parallel transport of
the entire path ${\tilde \gamma}$.\\

\medskip


\underline{\textit {Categorical picture}}

\begin{figure}[ht]
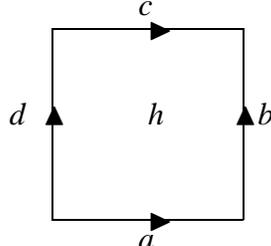

\setlength{\unitlength}{.5in}
\plaq{a}{b}{c}{d}{h}
   \caption{Plaquette}
   \label{fig:Plaquette}
\end{figure}


Now consider a plaquette as in Figure \ref{fig:Plaquette},
whose edges are labeled with the elements of the group $G$. Here
${a}$ and $c$ are $\bar A$-parallel transports, $d$ and $b$ are
$A$-parallel transports, and   $h \in H$, all related by 
\begin{equation}
\tau(h)={a}^{-1}  b^{-1}  c  d \label{paralleltransportofapath} 
\end{equation}
The above equation leads to the $2$-categorical picture, where
the set of objects for both of the categories is the group $G$ and
the set of morphisms is $G^4 \times H$, in the above figure the
morphism is $(a, b,c,d\,; h)$, for the
category {\bf Vert} (vertical category) and {\bf Horz} (horizontal
category), the {\em source} and {\em targets} are as follows
\begin{eqnarray}
&&s_{\bf Vert}(a, b,c,d\,; h)=a\nonumber\\ 
&&t_{\bf Vert}(a, b,c,d\,; h)=c\nonumber\\ 
&&s_{\bf Horz}(a, b,c,d\,; h)=d\nonumber\\ 
&&t_{\bf Horz}(a, b,c,d\,; h)=b\nonumber
\end{eqnarray}
Keeping the condition \eqref{paralleltransportofapath} in mind, the composition law for
{\bf Vert} is given by 
\begin{equation}
(a, b,c,d\,; h)\times (c, b',d,d'\,; h')=\bigl(a,  b'   b, d, d'  d\,; h(\alpha
(d)h')\bigr)\label{verticalcomposition}  
\end{equation}
and that of {\bf Horz} is given by
\begin{equation}
(a, b,c,d\,; h)\bullet (a', f,c',b\,; h')=\bigl(a'  a, f, c'  c, d\,;(\alpha(d^{-1})h')
h\bigr)\label{horizontalcomposition}  
\end{equation}
It is easy to check that the identity morphism $a \rightarrow a$
for {\bf Vert} is $(a, e, a,e\,;e)$ and inverse of $(a, b,c,d\,; h)$ is
$(c, b^{-1},a, d^{-1}\,;\alpha(d)h^{-1})$, on the other hand for the
{\bf Horz} the identity morphism $d\rightarrow d$ is $(e,d,e,d\,;e)$
and the inverse of $(a, b,c,d\,; h)$ is $(a^{-1}, d,c^{-1},b\,; \alpha
(a)h^{-1})$, here $e$ denotes the identity element for both $G$ and $H$. The 
category axioms can be readily verified.


In ordinary gauge theory a parallel transport operator ${\mathcal
  H}(\gamma, 0, 1)$ between $\gamma (0)$ and $\gamma (1)$ along the
path $\gamma$ transforms homogeneously as $U(\gamma (1)) {\mathcal
  H}(\gamma, 0, 1) U(\gamma(0))^{-1}$, here $U(\gamma (0))$ and
$U(\gamma (1))$ are two elements of the gauge group associated with
the end points of the path and ${\mathcal H}(\gamma, 0, 1)$ is also
an element of the same group. Now consider a plaquette as in
Figure \ref{fig:Plaquette}. Here instead of a group-valued parallel
transport operator we have a morphism like $(a, b, c, d\,;h)$ and
have two end paths rather than two end points. So in the same
spirit we define gauge transformation of a surface parallel
transport operator as
\begin{equation}
  (\bar a,\bar b,\bar c,\bar d\,;\bar h)\stackrel{\rm def}{=}(c,
  {\tilde U}(1), c, {\tilde U}(0)\,;\tilde W )\times
  (a,b,c,d\,;h)\times (a, U(1), a, U(0)\,;W 
  )^{-1} \label{gaugetransformation} 
\end{equation}  
Here $U(0), U(1)\in G$ are group elements associated with the left
and right end points of the initial path in
Figure \ref{fig:Plaquette} respectively, ${\tilde U}(0), {\tilde
  U}(1)\in G$ are those of the final path, and $W,{\tilde W} \in H$
are path ordered exponentials of some $LH$-valued one form
$\lambda$ over the initial and the final path respectively. As we
have already defined the vertical composition in
\eqref{verticalcomposition}, from \eqref{gaugetransformation} we
have following transformations
\begin{eqnarray}
&&\bar a = U(1)\cdot a\cdot  \tau (W)\cdot  U(0)^{-1}\nonumber\\
&&\bar b ={\tilde U}(0)\cdot b\cdot U(0)^{-1}\nonumber\\
&&\bar c={\tilde U}(1)\cdot c\cdot \tau ({\tilde W})\cdot  {\tilde U}(0)^{-1}\nonumber\\
&&\bar d={\tilde U}(1)\cdot b\cdot U(1)^{-1}\nonumber\\
&&\bar h=(\alpha(U(0))(W^{-1}.h.((\alpha(d^{-1}){\tilde W}))\nonumber
\end{eqnarray}

To conclude, we summarize the main points: (i) we have described how connections $A$, $\bar A$ on a bundle, and a $2$-form $B$ taking values
in a different Lie algebra, give rise to a connection over path-spaces, (ii) we described a pair of categories which arise from considerations
of parallel-transport along paths and surfaces, (iii) we outlined ideas on the effect of gauge-transformations on the categorical/parallel-transport structures.

\medskip

{\bf Acknowledgments} 
 ANS acknowledges research supported
from US NSF grant DMS-0601141. AL acknowledges research support from
Department of Science and Technology, India under Project
No.~SR/S2/HEP-0006/2008.


\bibliographystyle{aipproc}   



\end{document}